\documentclass[twocolumn]{assets/apie}
\pdfoutput=1
\usepackage{cite}
\makeatletter
\newcommand{\removelatexerror}{\let\@latex@error\@gobble}
\makeatother
\usepackage{amsmath,amssymb,amsfonts}
\usepackage{graphicx}
\usepackage{caption}
\usepackage{subcaption}
\usepackage{textcomp}
\usepackage{xcolor}

\usepackage{algorithmic}
\usepackage{bm}
\usepackage{diagbox}
\usepackage{booktabs}
\usepackage{url}
\usepackage{multirow}
\usepackage{geometry}
\newgeometry{
  top=72pt,
  bottom = 54pt,
  left = 54pt,
  right = 54pt
}

\usepackage{algorithm}
\usepackage{algorithmic}
\usepackage{array}
\usepackage{textcomp}
\usepackage{stfloats}
\usepackage{url}
\usepackage{diagbox}
\usepackage{verbatim}
\usepackage{graphicx}
\usepackage{multirow}
\usepackage{cite}
\usepackage{xcolor}
\usepackage{amsthm}

\usepackage{stfloats}

\usepackage{tikz}
\usetikzlibrary{positioning,arrows.meta} 
\usepackage{graphicx}

\usepackage{bbm}

\def\BibTeX{{\rm B\kern-.05em{\sc i\kern-.025em b}\kern-.08em
    T\kern-.1667em\lower.7ex\hbox{E}\kern-.125emX}}

\usepackage{microtype}
\usepackage{float}

\title{To Defer or To Shift? The Role of AI Data Center Flexibility on Grid Interconnection}

\author[1]{Yize Chen}
\author[1]{Xiaogui Zheng}

\affiliation[1]{University of Alberta}

\definecolor{verylightgray}{rgb}{0.9,0.9,0.9}
\definecolor{lightblue}{rgb}{0.733,0.875,1.0}
\definecolor{metablue}{rgb}{0, 0.392, 0.898}
\definecolor{forestgreen}{rgb}{0.208,0.667,0.235}
\definecolor{verylightblue}{rgb}{0.839,0.925,1.0}
\definecolor{veryverylightblue}{rgb}{0.92,0.96,1.0}

\abstract{
The integration of AI data centers into power grid represents one of the most emerging and complex challenges for the energy systems. As computational demand scales at an unprecedented rate, the traditional grid planning study's paradigm of treating data centers as rigid, inflexible loads is becoming economically, mathematically and operationally untenable. This work tries to understand and address the large load  interconnection bottleneck by modeling and evaluating AI load flexibility. By examining data center's temporal and spatial shifting capabilities within a grid capacity expansion framework, we build a quantitative grid planning model, and evaluate their impacts on additional generation, operational costs, and network congestion. Numerical study reveals interesting observations, as AI data center flexibility are not felt consistently, and increasing flexibility does not necessarily translate to less generation capacity required. Depending on data center's locations, flexibility range, and grid load conditions, flexible AI load can help reduce grid investment and operational costs by $3-21\%$. Our work also indicate that longer deferral time of AI compute has diminishing returns for offloading grid electricity dispatch pressure. }

\date{\today}
\correspondence{Yize Chen \email{yize.chen@ualberta.ca}}

\begin{document}

\maketitle

\section{Introduction}

The rapid growth of Generative Artificial Intelligence (GenAI) and large language models (LLM) has fundamentally altered the trajectory of long-term utility load forecasting. By the year 2035, independent projections estimate that power demand from AI data centers (AIDCs) in the United States alone could grow more than 30-fold, reaching a staggering 123 GW of demand, up from a baseline of just 4 GW in 2024~\citep{deloitte_ai_infrastructure_2025}. These large load interconnection requests have completely overwhelmed traditional grid capacity expansion pipelines. The median time required from a preliminary interconnection request to commercial operation has steadily increased to over five years approximately.

Today's AI data centers (DC) are mostly served under a firm service model, where the grid must be able to provide whatever power these facilities need at any time. Both AI model training and inference are highly resource-intensive, which requires sustained, burst-heavy compute cycles utilizing massive clusters of highly interconnected GPUs or TPUs. Maintaining satisfiable level of service requires significant investment in generation and infrastructures such as new and upgraded power plants, balancing stochastic demand and generation from time to time, raising costs that ultimately get passed down to both data center users and electricity consumers~\citep{li2024unseen}. 

\begin{figure}[t]
    \centering
    \includegraphics[width=0.5\textwidth]{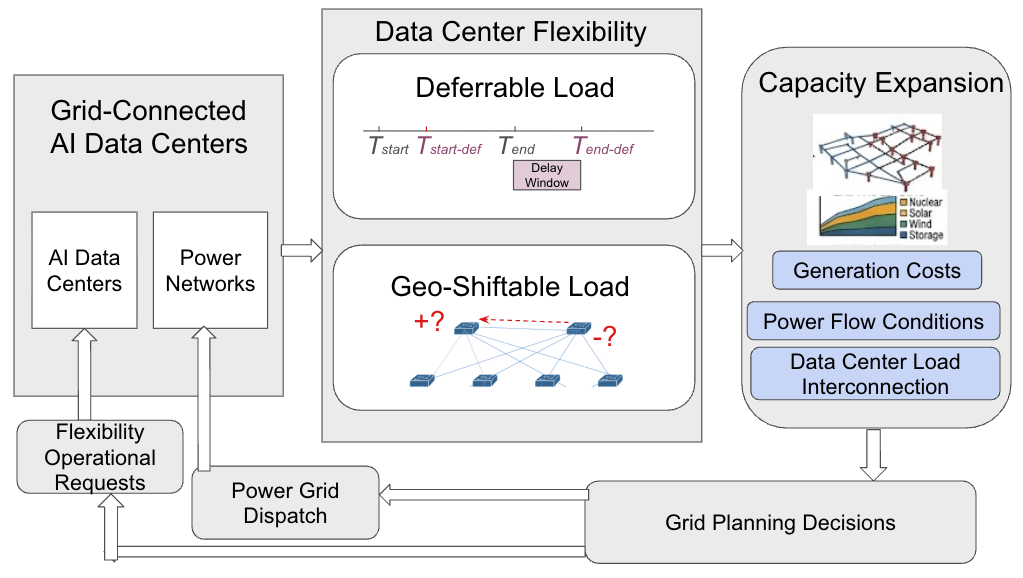}
    \caption{This work studies how data center flexibility, namely temporally deferrable load and geographically shiftable load, would impact grid capacity expansion decision. We also ensure grid planning decisions are feasible to serve power flow and flexible DC demands.}
    \label{fig:schematic}
\end{figure}

To avoid such an economical challenge and help accelerate AI data centers interconnection, flexible data centers have gained traction as a promising approach.  Unprecedented AI advancements and connection delays have recently led to growing discussions among data-center-side solutions~\citep{frick2025large}, such as load curtailment~\citep{zheng2020mitigating}, installing energy storage and flywheels~\citep{yan2026switching}, and colocating with existing generation facilities~\citep{guo2017colocation}. Among them, AI load flexibility can offer a practical route for not only accelerating the interconnection of large demand loads, but improving power grid reliability and reducing system costs under future uncertain renewable generations or extreme events. Rather than provisioning for the absolute peak demand of every DC, by utilizing load flexibility using techniques such as throttling or delaying usage, shifting load to other data centers, it is possible for electrical grid to reduce coincident peak and properly manage grid congestion~\citep{liu2013data, lindberg2021guide}. Recent prototype studies on real GPUs and power grids also show early promises of such a flexibility scheme in   mitigating the pressure of rapid resource and transmission expansion~\citep{nationalgrid_emeraldai_2025}. More hyperscalers and data center developers work with utilities and grid operators, and start to take actions to mitigate grid interconnection pressure by prioritizing near-term solutions for load flexibility~\citep{google_demand_response_2026}.

To that end, DC can be treated as a provider for demand response, so that when grid operators implement power dispatch, peak demand can be largely curtailed, shifted to non-congested regions, or delayed to later times. While it sounds viable, to date the exact flexibility which could be supplied by AI loads are unclear. Moreover, utilities, system operators, and policymakers are still using firm load model when running grid planning studies. More importantly, regarding the specific benefits and conditions of AI load flexibility, we are interested in evaluating the possible routes for the following question:

\emph{Given large load interconnection targets, what kind of and how much data center flexibility is needed?}

In this work, we consider the setting of grid capacity expansion with growing penetration of AI data center load, where capacity addition decisions are introduced to achieve both long-term resource adequacy and short-term grid operational reliability. To examine load flexibility's role in impacting grid expansion decisions, we propose to integrate two types of AI compute flexibility convertible to data center demand flexibility: \emph{deferrable workload} and \emph{geographically shiftable workload}. By formulating a two-stage capacity expansion model with grid constraints and AI load serving requirements, our work connects data center flexibility with grid interconnection and capacity expansion problem in a quantitative approach. See Fig. \ref{fig:schematic} for the workflow.

Surprisingly enough, implications of this study are somewhat different from current literature, which largely omits capacity planning optimization or network flow models~\citep{norris2025rethinking}. For numerical experiments with goals to minimize grid investment and operational costs on IEEE 14-bus and WECC 240-bus systems~\citep{price2011reduced}, there are interesting scenarios where data center (DC) flexibility indeed \emph{increase} the need of future generation capacity. This could happen because flexibility can increase utilization of the system in a way that makes expansion more valuable. Without flexibility, some DC demand may be stuck at expensive or weak-connected locations, so the model operates conservatively. With DC flexibility mechanism, the model may discover a lower-cost operating regime that is only worthwhile if a bit more capacity is added. This reveals flexibility improve grid congestion management, which in turn enables cheaper generation usage and also increases marginal value of capacity. 

Overall, both geographical and temporal flexibility can help reduce grid operational costs ranging from 3\% to more than 10\%. While our work also shows that not all flexibility are equally important: the benefits on improving grid conditions diminish significantly when load is deferred for more than 3 hours. While for geographical load shifting, the benefits brought to cost savings highly depend on grid congestion level. In general, additional geographical shifts are more helpful when powerlines are congested, therefore load can be redistributed and served by regions with larger headrooms.

\subsection{Related Works}

\subsubsection*{Hosting Capacity Analysis and Grid Interconnection Study} With the everchanging landscape of renewable generation and electrification, there is a significant body of literature on enhancing power grid planning by integrating distributed energy resources~\citep{ranjbar2021resiliency}, demand response~\citep{anjo2018modeling}, and energy storage~\citep{xiong2015optimal}, to improve grid resilience and reliability~\citep{dharmasena2022algorithmic, georgilakis2015review}. 
To evaluate and decide grid and generation expansion plan, demand curves can be built using Monte Carlo tools which simulate outage, load, and renewable uncertainty~\citep{stenclik2020redefining}. With growing AI compute, recent works also explicitly incorporate the siting of large-scale flexible loads into stochastic nodal power system capacity expansion models. In these frameworks, the placement and sizing of large loads are co-optimized concurrently with traditional decisions regarding generation construction, transmission line expansion, and energy storage deployment~\citep{zuluaga2025nodal}. In \citep{gu2025role}, two flexibility types, the delay and curtailment flexibility are examined under the setting of distribution grid interconnection while limiting the number of utility-controlled interventions. While how flexibility modeling and flexibility level would impact capacity expansion and DC interconnection capacity performance is underinvestigated.

\subsubsection*{Data Center Flexibility} Previous works have considered demand response of computational load in data centers, mostly using CPU-based clusters running high-performance computing (HPC) applications~\citep{zhang2021data, aljbour2024powering}. These studies have revealed valuable insights around DC load flexibility, but have largely not accounted for service level objectives (SLOs) of AI load and distinct energy profiles of AI training and inference workloads on GPUs. Studies also indicate the potential of spatiotemporal DC load shifting for carbon emission and energy reductions~\citep{radovanovic2022carbon, he2025freesh}, and serving for grid demand response objectives~\citep{mehra2023supporting}. \citep{crozier2025potential} describes types of DC flexibility; and \citep{caprara5385595flexibility} quantifies the deferral capabilities of DC workloads based on real industrial workload trials Recent field study also demonstrate data centers operating as flexible grid resources and reducing power usage by 25\% for 3 hours during peak demand while maintaining AI quality of service guarantees \citep{colangelo2025ai}. One study suggests that U.S. grid may have more than 100 GWs of headroom for new load without the need for major capacity expansion~\citep{norris2025rethinking}, while it is based on a coarse-grained AI load flexibility model without considering network power flow constraints.

\section{Capacity Expansion Model with Data Centers}

\subsection{Data Center with Flexible Load}
Since in practice, it is viable to convert AI serving SLOs to compute need in terms of energy usage or computation time, we can translate DC workload computing constraints to either temporal deferral constraints or location power shift constraints. For instance, a LLM inference SLO might mandate that $99\%$ of call requests must be completed within a certain latency threshold, while a bulk processing pipeline such as model training/finetuning might have a throughput SLO demanding a specific volume of gigabytes processed per hour. 

To model these loads' interconnection in grid capacity expansion problem, we consider a power network with heterogeneous demand components, including nominal (inelastic) load, as well as temporally deferrable and geographically shiftable DC load. Let $\mathbf{L}_t, \mathbf{L}_t^{\text{base}}, \mathbf{L_t^{\text{def}}}, \mathbf{L}_t^{\text{geo}}$ denote the nodal aggregate, nominal, temporally deferrable, and geographically shiftable  load vector at time $t$, which can be decomposed as
\begin{equation}
\mathbf{L}_t = \mathbf{L}_t^{\text{base}} + \mathbf{L}_t^{\text{def}} + \mathbf{L}_t^{\text{geo}},
\end{equation}

\paragraph{Deferrable workload model}
For many AI training and inference jobs, they have a flexible window so that users could opt for later completion time. In practice, it is possible to determine the delay window size based on user tolerance for runtime or throughput deviations. Mathematically, let $\mathbf{u}_t$ denote the arriving deferrable workloads, $\mathbf{s}_t$ the served workloads across deferrable classes, and $\mathbf{b}_t$ the backlog state for unserved workloads. Note that $\mathbf{u}_t, \mathbf{b}_t, \mathbf{s}_t$ have the unit of $MW$, as in this work we assume all AI compute can be converted to energy consumption requests, which is shown to be practical for modern LLM loads~\citep{he2025freesh, stojkovic2025dynamollm}. Then these states evolve as
\begin{subequations}
\label{equ:backlog} 
\begin{align}
& \mathbf{b}_t = \mathbf{b}_{t-1} + \mathbf{u}_t - \mathbf{s}_t,\\
& \mathbf{0} \le \mathbf{s}_t \le \mathbf{b}_{t-1} + \mathbf{u}_t,  \\
& \mathbf{b}_t \ge \mathbf{0},
\end{align}
\end{subequations}
with $\mathbf{b}_0=\mathbf{0}$ and $\mathbf{b}_T=\mathbf{0}$. Moreover, for each delayable class $d$ with delay window $W_d$, the following condition is required to ensure job completion
\begin{align}
\sum_{\tau=1}^{t} s_{d,\tau}
\ge
\sum_{\tau=1}^{t-W_d} u_{d,\tau},
\qquad
t=W_d+1,\dots,T. \label{equ:temporal_logic}
\end{align}

The realized deferrable load injected into the network is mapped via
\begin{equation}
\mathbf{L}_t^{\text{def}} = A^{\text{def}} \mathbf{s}_t, \label{equ:def_map}
\end{equation}
where $A^{\text{def}}$ maps workloads to buses.

\paragraph{Geographically shiftable demand model.}
Let $\mathbf{L}_t^{\mathrm{geo}, 0}$ denote the baseline geographically shiftable demand at time $t$, and let $\boldsymbol{\delta}_t \in \mathbb{R}^N$ denote the demand-shifting decision. The realized geographical demand is
\begin{align}
{\mathbf{L}}_t^{\mathrm{geo}} = \mathbf{L}_t^{\mathrm{geo}, 0} + \boldsymbol{\delta}_t. \label{equ:geo}
\end{align}

The shifting vector satisfies
\begin{subequations}
\label{equ:geoshift}
\begin{align}
& \mathbf{1}^\top \boldsymbol{\delta}_t = 0, \\
& \underline{\boldsymbol{\delta}}_t \le \boldsymbol{\delta}_t \le \bar{\boldsymbol{\delta}}_t,
\end{align}
\end{subequations}
which ensures that total geographically shiftable demand is conserved, and that each bus respects its shifting bounds. Optionally, a total shifting budget may be imposed as
\begin{align}
\|\boldsymbol{\delta}_t\|_1 \le \Gamma_t.\label{equ:allshift}
\end{align}

\subsection{Capacity Expansion With Deferrable and Geographically Shiftable Data Center Loads}

We consider a transmission network represented by a graph $\mathcal{H} = (\mathcal{N}, \mathcal{E})$, where $\mathcal{N}$ is the set of buses and $\mathcal{E}$ is the set of transmission lines. Power flows are modeled using the DC approximation. Let $\mathbf{p}_t \in \mathbb{R}^{|\mathcal{G}|}$ denote generation dispatch, $\mathbf{\theta}_t \in \mathbb{R}^{|\mathcal{N}|}$ the voltage phase angles, and $\mathbf{f}_t \in \mathbb{R}^{|\mathcal{E}|}$ the line flows.  A reference bus is fixed by setting $\theta_{\text{ref},t} = 0$. The load  vector $\mathbf{L}_t$ integrates nominal demand and flexible DC loads, thereby coupling network feasibility with spatio-temporal load shifting decisions.

We formulate a two-stage capacity expansion problem over a time horizon $\mathcal{T}=\{1,\dots,T\}$ with hourly resolution. The first stage determines generation expansion decisions, while the second stage optimizes hourly dispatch and flexible DC demand.

\paragraph{First stage}
Let $x_g$ denote the added capacity of candidate generator $g\in\mathcal{G}$, and let $C_g^{\mathrm{inv}}$ denote the corresponding investment cost. Denote the operational costs from the second stage as $Q(\mathbf{x})$, which will be detailed later. The first-stage problem is a planning setting, which can be formulated as
\begin{align}
\min_{\mathbf{x}} \quad
& \sum_{g\in\mathcal{G}} C_g^{\mathrm{inv}} x_g + Q(\mathbf{x}) \\
\text{s.t.}\quad
& 0 \le x_g \le \bar{x}_g, \qquad \forall g\in\mathcal{G}.
\end{align}

\paragraph{Second stage}
Given the grid expansion decision $\mathbf{x}$, the second-stage operating problem involving served DC loads $\mathbf{s}_t$, geographically shiftable load $\mathbf{L}_t^{\mathrm{geo}}$, and nominal loads $\mathbf{L}^{\mathrm{base}}_t$ is
\begin{subequations}
\begin{align}
Q(\mathbf{x}):=
\min_{\substack{\mathbf{p}_t,\boldsymbol{\theta}_t,\mathbf{f}_t,\\
\mathbf{s}_t,\mathbf{b}_t,\boldsymbol{\delta}_t}}
\quad
& \sum_{t\in\mathcal{T}}
\left(
\mathbf{c}^\top \mathbf{p}_t
+ (\boldsymbol{\pi}^{\mathrm{delay}})^\top \mathbf{b}_t
+ \pi^{\mathrm{shift}} \|\boldsymbol{\delta}_t\|_1
\right)
\end{align}
\end{subequations}
subject to, for all $t\in\mathcal{T}$,
\begin{subequations}
\begin{align}
& \mathbf{0} \le \mathbf{p}_t \le \bar{\mathbf{P}}^0 + \mathbf{x}, \\
& \mathbf{f}_t = M \boldsymbol{\theta}_t, \\
& -\bar{\mathbf{f}} \le \mathbf{f}_t \le \bar{\mathbf{f}}, \\
& A^{\mathrm{gen}}\mathbf{p}_t - \mathbf{L}^{\mathrm{base}}_t - A^{\mathrm{def}}\mathbf{s}_t - \mathbf{L}^{\mathrm{geo}}_t = K \mathbf{f}_t, \\
& \theta_{\text{ref},t}=0,\\
& \eqref{equ:backlog} \, \eqref{equ:temporal_logic}\, \eqref{equ:def_map} \, \eqref{equ:geo}\, \eqref{equ:geoshift}\, \eqref{equ:allshift}
\end{align}
\end{subequations}

Here, $\mathbf{c}$ is the generation cost vector, $A^{\mathrm{gen}}\in\{0,1\}^{|\mathcal N|\times |\mathcal G|}$ is the generator-to-bus mapping matrix, $M\in\mathbb{R}^{|\mathcal E|\times |\mathcal N|}$ is the DC flow operator, and $K\in\mathbb{R}^{|\mathcal N|\times |\mathcal E|}$ is the bus-line incidence matrix. The backlog penalty $\boldsymbol{\pi}^{\mathrm{delay}}$ models accumulated delay cost, analogous to holding cost in inventory systems. Also note that $\mathbf{L}_t$ is partially decision-dependent through $\mathbf{s}_t$ and $\boldsymbol{\delta}_t$, which distinguishes flexible demand from exogenous base load.

To solve this problem on power and compute data, we start with an auxillary load shedding cost variable to find feasible solutions on capacity expansion, and then gradually move out load shedding to find feasible range. We also vary the parameters of $W_d$ and $\boldsymbol{\delta}$ to evaluate how DC flexibility construction would impact the generation expansion decisions.

This capacity planning model considers hourly compute load, and identifies how much generation or network capacity must be added to reliably serve growing DC demand, and how those requirements change when a portion of the load can be shifted across locations or deferred in time. This allows planners, utilities, and large-load customers to move beyond qualitative claims about flexibility and instead determine the specific flexibility level needed to avoid costly upgrades, relieve local congestion, or accelerate interconnection timelines. Alternatively, it could also help examine the project practicability given a budget of investments or flexibility characterization. More interestingly, it could offer a quantitative decision-support tool for evaluating whether it is more cost-effective to expand physical infrastructure or to procure and coordinate flexible computing demand.

Although current deterministic capacity expansion remains convex with respect to grid expansion decision and operating decisions, its size grows rapidly with the number of buses, generators, candidate investments, and time periods, since the second-stage recourse variables must be instantiated at every hour, and the temporal constraints \eqref{equ:backlog} defined over deferrable load backlog state make each timestep non-separable. For larger transmission systems or richer first-stage design spaces (e.g., simultaneous generation, transmission, storage, and DC siting decisions), solving the full deterministic equivalent monolithically may become computationally burdensome. In such settings, decomposition methods such as Benders decomposition becomes attractive, which will be left as future work. Representative-period or scenario reduction methods can also reduce the recourse dimension while preserving key seasonal and diurnal patterns.

\section{Simulation Setup}
To quantitatively evaluate the planning value of DC flexibility, we implement the proposed model on both the IEEE 14-bus system and the WECC 240-bus system. The IEEE 14-bus case is used as a transparent small-scale benchmark for validating the behavior of the flexibility model and visualizing geographical load shifting behaviors. The WECC 240-bus case is used to examine the planning implications of flexible DC demand in a more realistic transmission system with heterogeneous load and network congestion patterns.

\begin{figure}[t]
    \centering
    \includegraphics[width=0.5\textwidth]{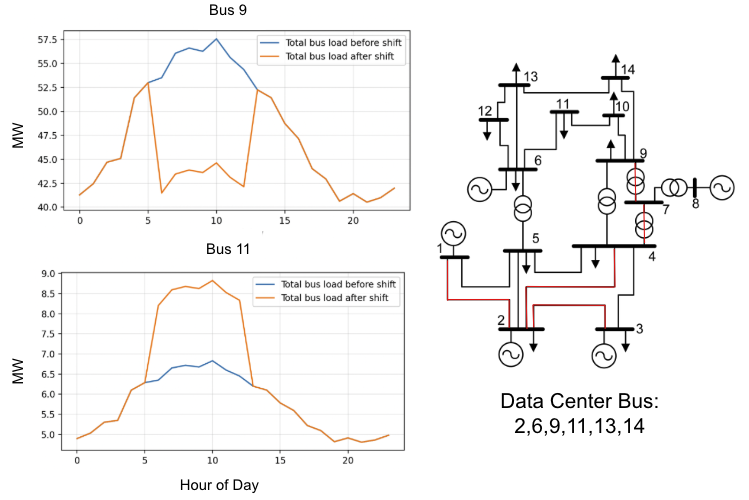}
    \caption{A 24-hour snapshot of geographical AI load shifting on IEEE 14-bus case for two buses with DC load (left), where congested lines are marked in red (right).}
    \label{fig:daily-profile}
\end{figure}

For each system, we construct an hourly load profile over the planning horizon (full-year data with 30 days for representative simulation) based on WECC's load profiles~\citep{wecc_loads_resources_2026}, and superimpose DC demand on a subset of buses. The data center load is modeled as a combination of  temporally deferrable demand and geographically shiftable demand. In the geographical shifting setting, only a portion of the nodal DC load is designated as flexible and can be reallocated among selected buses subject to conservation and bus-level shifting limits. In the temporal shifting setting, randomly selected nodes are  with deferrable workloads, which are represented through backlog dynamics and completion-window constraints. Throughout the numerical study, we are interested in answering the following questions:

\begin{itemize}
    \item Compared to the scenario without data center flexibility, how much cost and capacity can be improved?
    \item With growing penetration of data center, how quickly cost escalates, and whether flexibility delays the onset of expensive upgrades?
    \item Given limited power system expansion budget, how much flexibility must the data center provide?
\end{itemize}

We consider several groups of experiments. First, we vary the flexibility level, including the delay window size $W_d$ and the proportion of geographically flexible data center load, to study their impact on total cost, investment cost, and required added capacity. Second, we vary data center load penetration to characterize how rapidly system cost and expansion needs increase as large load interconnection grows. Third, for a given expansion budget or target planning cost, we determine the minimum flexibility needed to maintain feasibility or achieve the desired interconnection target. Across all experiments, we record total system cost, investment cost, added generation capacity, shifted-load magnitude, and congestion indicators such as the number of binding transmission lines and peak generators.

\section{Scalability Evaluation}

\begin{figure}[t]
    \centering
    \includegraphics[width=0.5\textwidth]{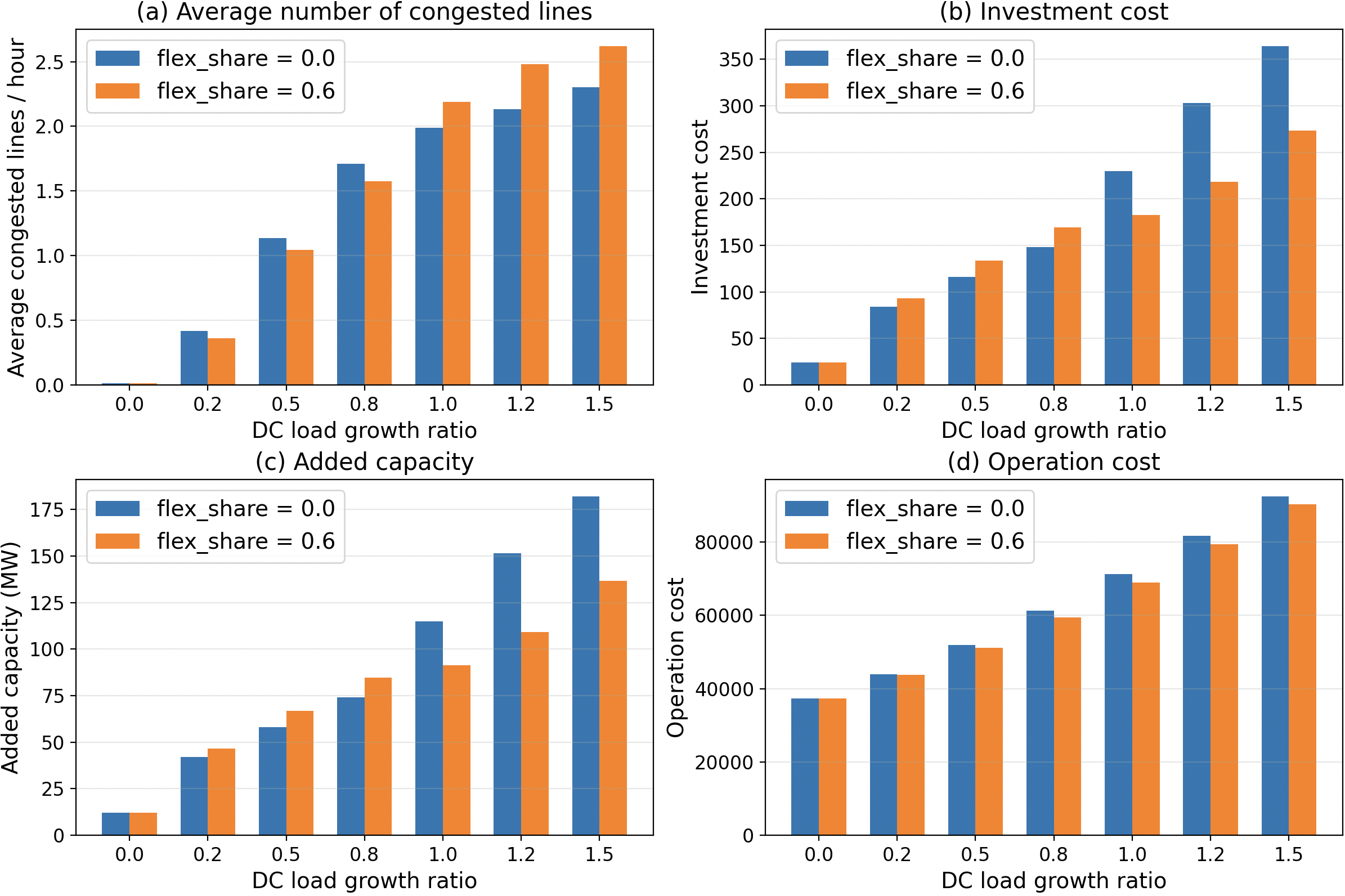}
    \caption{Effects of load growth on capacity expansion with/without geographical load shifting. Simulations are conducted on IEEE 14-bus system with 1.0x (right) line capacity. DC load growth ratio is defined over DC load versus nominal nodal load for each bus respectively.}
    \label{fig:flex-growth}
\end{figure}

\begin{figure*}[t]
    \centering
    \includegraphics[width=\textwidth]{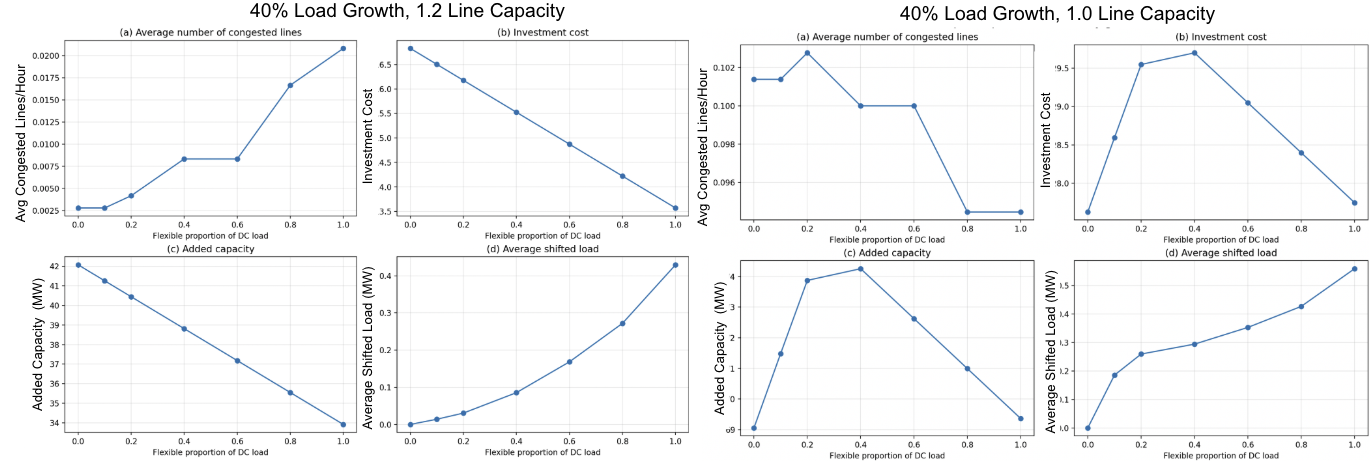}
    \caption{Effects on capacity expansion with varying level of flexible load participation. Simulations are conducted on IEEE 14-bus system with 1.2x (left) and 1.0x (right) line capacity.}
    \label{fig:flex-portion}
\end{figure*}

In Fig. \ref{fig:daily-profile} we illustrate a daily operational profile for the geographically shiftable workload, along with the IEEE 14-bus network condition, with binding lines marked in red. It is interesting to observe that during peak hours, DC buses connected to congested lines (e.g., Bus 9) are shifting their load to non-congested regions, thereby the network is able to sustain heavier load conditions. This indicates that one role of geographical AI load shifting is to help mitigate line congestion.

We also evaluate how load flexibility could aid under varying DC load penetration. Fig. \ref{fig:flex-growth} compares results without flexible load and with 60\% of the load geographically shiftable. It indicates that when load are relatively small, flexible loads mainly help for reducing operational costs, and additional capacity could help shift dispatch to lower-cost generations. While when AI load become more significant in the network, it also helps reduce generation capacity investment. Under the most extreme scenario where we test the DC load is 1.5x of nominal non-DC load, the flexible AI data centers help reduce more than 21\% of the generator capacity.

To examine how geographical shifting's flexibility level would impact the grid expansion decision, in Fig. \ref{fig:flex-portion} we vary the proportion of AI load which are shiftable across DCs and investigate its impacts. The results are quite diverging giving different load conditions. When lines are relatively non-binding (1.2x Line Capacity), increasing load flexibility can monotonically reduce capacity investment by over 21\%. While the situation is more complex when networks are more congested (1.0x Line Capacity), and resulting capacity addition could either increase or decrease, depending on flexible load's participation rate. It is interesting to observe cases where flexibility can reduce operating cost enough that the planning model is willing to spend more on generation capacity expansion overall to unlock even larger operating savings~(Fig. \ref{fig:flex-portion}).  So total cost still goes down by 3.5\% even though investment goes up by 7.1\%.

\begin{figure}[t]
    \centering
    \includegraphics[width=0.5\textwidth]{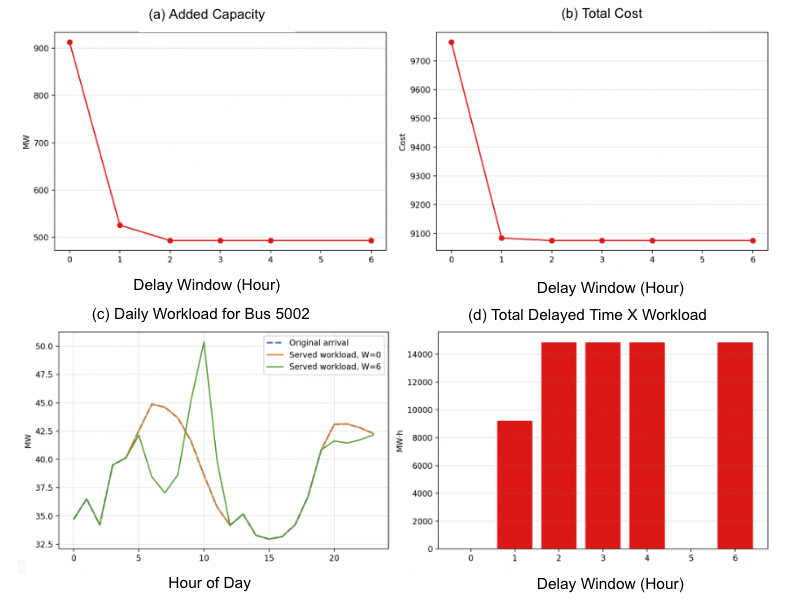}
    \caption{Results on WECC 240-bus test system with AI DC load of varying deferrable window. }
    \label{fig:temporal}
\end{figure}

Fig. \ref{fig:temporal} shows results of deferrable load on WECC 240-bus system. With temporally delayable work, the AI load can be shifted to non-peak hours. But the delay's benefits towards network operational and planning costs are almost the same when delay window is greater or equal to 3 hours. This suggests that when calling for flexibility from data centers, short-term flexibility is already providing necessary benefits for grid tasks. 

It is worthwhile to note that flexibility does not come without a price. There are tradeoffs inherent in certain demand response programs, such as increasing uncertainty, extra procedural and operational requirements, and additional costs for future data center projects left in the interconnection queue. While this study mainly takes the power grid perspective.

\section{Conclusion and Future Works}

In this work, we demonstrate the varying potentials of data centers load flexibility for power grid interconnection study. By imposing network flow constraints and AI flexibility requirements into the capacity expansion framework, we are able to evaluate the interplay between load growth, data center flexibility, and system costs. Our simulation study reveals surprising observations, indicating that while flexible AI load help reduce system costs as a whole, it does not necessarily reduce additional generation capacity. Our work also characterize how AI load's deferral duration and flexibility participation rate would impact the generation interconnection. 

In practice, choosing the optimal expansion decision of capacity and computing is challenging. In particular, designing future grids involves choosing which power system and data center components to invest (e.g., among generators, energy storage, new transmission lines, GPU models, cooling systems, data center configurations) at what size and in which location. This requires evaluating thousands of binary investment decisions by representing such decision in a realistic transmission grid which may involve thousands of buses. In addition, for modern AI load, model training and model inference possess fundamentally different hardware architectures, power draw profiles, latency sensitivities, and grid flexibility potentials, which will be examined in more details for capacity expansion studies. Load especially AI load's uncertainty will also need to be taken into account in a two-dtage stochastic programming framework in the future work.

\bibliographystyle{abbrvnat}
\bibliography{bib}

\end{document}